\documentclass[a4paper,11pt]{article}
\usepackage{jheppub} 
\usepackage{lineno}

\usepackage{diagbox}

\usepackage{mathtools}
\usepackage{amsfonts}
\usepackage{mathrsfs}
\usepackage{bbm}
\usepackage{slashed}
\usepackage{dsfont}

\usepackage{graphicx}
\usepackage{color}
\usepackage{array}

\usepackage{blindtext}
\usepackage{placeins}
\usepackage{booktabs}
\usepackage{makecell}
\usepackage{epstopdf}
\usepackage[caption=false]{subfig}

\usepackage{xspace}
\usepackage{siunitx}
\usepackage{hyperref}
\usepackage[nameinlink]{cleveref}
\usepackage{appendix}

\usepackage{xifthen}
\usepackage{xcolor}
\hypersetup{
	colorlinks,
	linkcolor={red!75!black},
	citecolor={blue!75!black},
	urlcolor={blue!75!black}
}

\setkeys{Gin}{width=0.48\textwidth}

\graphicspath{
	{./figures/}
	{../figures/}
}

\def\s0#1#2{\mbox{\small{$ \frac{#1}{#2} $}}}
\def\0#1#2{\frac{#1}{#2}}

\def\Eq#1{Eq.~\eqref{#1}}

\newcommand{\imag}{\text{i}}

\renewcommand{\Re}{\operatorname{Re}}

\usepackage{multirow}
\usepackage{colortbl}
\definecolor{kugray5}{RGB}{224,224,224}
\newcommand{\PreserveBackslash}[1]{\let\temp=\\#1\let\\=\temp}
\newcolumntype{C}[1]{>{\PreserveBackslash\centering}p{#1}}
\newcolumntype{R}[1]{>{\PreserveBackslash\raggedleft}p{#1}}
\newcolumntype{L}[1]{>{\PreserveBackslash\raggedright}p{#1}}

\newcommand{\spatial}[1]{\boldsymbol{#1}}
\newcommand{\tr}{{\text{tr}}}

\title{Revisiting the first-order QCD phase transition in dense strong interaction matter}

\author[a]{Yi Lu,}
\author[b]{Fei Gao,}
\author[a,c,d]{Yu-xin Liu}

\emailAdd{qwertylou@pku.edu.cn}
\emailAdd{fei.gao@bit.edu.cn}
\emailAdd{yxliu@pku.edu.cn}

\affiliation[a]{Department of Physics and State Key Laboratory of Nuclear Physics and Technology, Peking University, Beijing 100871, China}
\affiliation[b]{School of Physics, Beijing Institute of Technology, 100081 Beijing, China}
\affiliation[c]{Center for High Energy Physics, Peking University, 100871 Beijing, China}
\affiliation[d]{Collaborative Innovation Center of Quantum Matter, Beijing 100871, China}

\abstract{We revisit the phase structure and thermodynamics of QCD in the low temperature and high density region, where a strong, first-order phase transition is expected beyond the critical end point. By solving the quark gap equation in the continuum QCD approach, we reveal the  coexistence  of the multi-phases both in the microscopic dynamics of chiral symmetry breaking and also in the thermodynamic observables, which manifests the existence of spinodal decomposition during the first-order QCD phase transitions. We also analyse the interface structure of the co-exist Nambu and Wigner phases in the isothermal process during the first-order transition. In particular, the interface tension and interface entropy density are extracted from the isothermal trajectories, which further allows for an analysis on the formation of nuclear bubble, including the bubble radius and its stability at different temperatures. Our predictions may serve as useful inputs for further investigations in heavy-ion physics or astrophysics research.}

\begin{document}
\maketitle
\flushbottom

\section{Introduction}

The thermodynamic properties of strong interaction matter is of great interest in nuclear and particle physics researches. 
In laboratory, the relativistic heavy-ion collision provides a systematic facility for experimentally probing the strong interaction matter and its phase transition at high and intermediate temperatures. 
In particular, the search of QCD critical end-point (CEP) signatures~\cite{STAR:2025zdq,STAR:2020tga,Bzdak:2019pkr,Luo:2017faz} is one of the main goals in future experiments at HIAF~\cite{Zhou:2022pxl}, FAIR~\cite{Klochkov:2021eyo} and NICA~\cite{Kekelidze:2017ual}. 
On the other hand, astrophysical observations are entering the multi-messenger era, with typical signatures from compact stars and their mergers which shed light on the equation of state of cold and dense strong interaction matter~\cite{LIGOScientific:2018cki,Annala:2017llu}. Studies have also indicated that a strong, first-order QCD phase transition can be responsible for the primordial gravitational wave signatures~\cite{Cline:2025bwe,Gao:2024fhm,Pasechnik:2023hwv}.
Besides, there are stimulating researches that connects the heavy-ion physics and astrophysics for a better understanding on the nuclear equation of state~\cite{Sorensen:2023zkk}. 

The investigation on these signatures requires a combined analysis between a great amount of experimental data and the theoretical predictions on the phase structure and the equation of state of QCD in precision. 
The latter however remains still as a long-standing problem due to the complicated nature of non-perturbative QCD. 
At zero baryon chemical potential $\mu_{B}^{}$, lattice QCD simulation as a first-principles approach has confirmed that the phase transition behaves as a crossover~\cite{Borsanyi:2020fev,HotQCD:2018pds}. 
The calculations on the equation of state have also been performed to the high orders of susceptibilities~\cite{Borsanyi:2023wno,Bazavov:2020bjn}. 
While at finite $\mu_{B}^{}$, lattice QCD is hampered from the sign problem and has to rely on extrapolations from the knowledge at zero $\mu_{B}^{}$~\cite{Borsanyi:2022qlh,Bollweg:2022fqq}. 
This makes the prediction controllable only within a small range of chemical potential, which is not yet possible for a direct access on the signatures of the conjectured CEP and for verifying those possible new physics beyond that, in particular the moat regime~\cite{Fu:2024rto}, inhomogeneous phases~\cite{Buballa:2014tba}, color superconductivity~\cite{Alford:2007xm} and so on. 
Aimed that the finite $\mu_{B}^{}$ region, theoretical approaches in the continuum space-time have been developed over the years, such as the chiral effective field theory~\cite{Drischler:2021kxf}, low energy effective models~\cite{Buballa:2008ru,Schaefer:2008ax}, holographic QCD~\cite{Rougemont:2023gfz,Chen:2022goa} and so on.
In particular, functional approaches including the Dyson-Schwinger equations (DSE) approach~\cite{Fischer:2018sdj} and the functional renormalisation group (fRG) approach~\cite{Fu:2022gou,Dupuis:2020fhh} of the continuum QCD, have the privilege as they are the only approaches to date that offer direct QCD computations at finite $\mu_{B}^{}$, with all required analytic properties of QCD as well as QCD dynamics extended to larger density~\cite{Aarts:2023vsf}. 

In this work, we revisit the QCD phase structure and thermodynamics of dense strong interaction matter using the DSEs approach. The applications of the DSEs approach with specific truncation schemes have covered the in-vacuum QCD dynamics~\cite{Ferreira:2023fva}, hadron spectrum~\cite{Ding:2022ows} as well as phase transitions at finite temperature and density~\cite{Gao:2020fbl,Gunkel:2021oya}. 
Specifically, we adopt a recently improved computational scheme from Refs.~\cite{Lu:2025cls,Gao:2020qsj} which takes into account both the confining and the chiral dynamics of QCD at finite density, whose prediction agrees on the lattice QCD benchmark results of thermodynamic functions at relatively small $\mu_{B}^{}$ and high temperature $T\gtrsim 100\,$MeV. 
We then directly extends the calculation to higher $\mu_{B}^{}$ and lower temperature region, where a first-order phase transition is expected~\cite{Gao:2020fbl,Gunkel:2021oya,Chen:2025ijh,Fu:2019hdw,Hippert:2023bel,Steinheimer:2013xxa,Wunderlich:2016aed,Pradeep:2024cca}. 
There, we find multi-phase coexistence in the chiral dynamics and also in the  thermodynamic quantities, implying that the spinodal decomposition, suggested by most low-energy effective models with the mean-field approximation, can be a genuine picture in first-order QCD phase transition. 
We further provide an estimate on the interface tension of nuclear bubbles, including their temperature dependence, based on an inhomogeneous configuration of nuclear density distribution calculated from the QCD isothermal trajectories during the first-order transition. 
The interface tension also allows us to analyse the stability of the formed nuclear bubbles through its compressibility at different temperatures. 

The paper is organised as follows: in \Cref{sec:phase-structure}, we illustrate the spinodal decomposition in the first-order QCD phase transition, which is observed both in the microscopic dynamics of quark propagator and also in the macroscopic observables - the chiral condensate and the Polyakov loop. 
In \Cref{sec:thermo} we investigate the thermodynamic properties and the equation of state of strong interaction matter at high density, where both the influence of QCD spinodal decomposition and the liquid-gas phase transition are taken into account. 
Then in \Cref{sec:interface}, we discuss the impact of spinodal decomposition on the interface effect, and provide an estimate on the interface tension and the formation of bubbles in dense strong interaction matter at different temperatures. 
Finally in \Cref{sec:summary}, we summarise the results and make some further outlooks.

\section{Phase structure in first-order QCD transitions}\label{sec:phase-structure}

Within the mean-field approximation, effective model studies have predicted the existence of spinodal decomposition in the QCD phase structure of first-order transitions~\cite{Xin:2014dia,Jiang:2013yoa}: the QCD effective potential $\Gamma$ as a function of the order parameter $\Delta$ shows two minimal points which correspond to the broken (Nambu) phase and the symmetric (Wigner) phase respectively; between these two phases, the spinodal region is found with the thermodynamic instability $\partial^2 \Gamma / \partial \Delta^2 < 0$, 
accompanied by the supercooling region and the superheating region. 
Going beyond the mean-field assumption, the spinodal region in non-perturbative QCD has also been indicated in a recent study via effective potential using the homotopy method~\cite{Zheng:2023tbv}: during first-order phase transition, an additional, local maximum is found in the effective potential between the $N$ (short for Nambu) and $W$ (short for Wigner) phases which typically shows thermodynamic instability. Here we would like to give a complete analysis within the up-to-date scheme in functional QCD approaches. 

\begin{figure}[htbp]
  \centering
  \includegraphics[width=0.7\columnwidth]{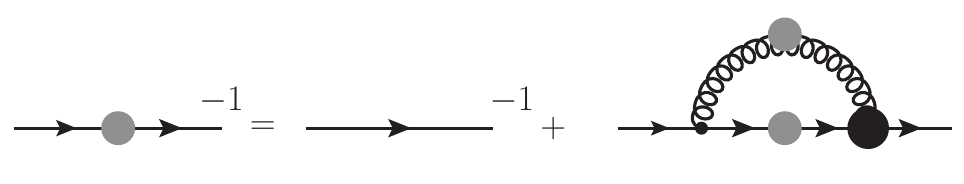}
  \caption{Diagrammatic description of the quark gap equation in QCD. The straight line with a gray blob is the full, non-perturbative quark propagator $G_q$ in \Eq{eq:quark-prop}, the curly line with a gray blob is the full gluon propagator, the black blob is the full quark-gluon interaction vertex, and the black dot is the classical quark gluon vertex.}\label{fig:quark-DSE}
\end{figure}

Via the quantum equation of motion of QCD, we verified that the spinodal decomposition can be a genuine phenomenon in the first-order QCD transitions. 
Particularly, the language of quantum equation of motion suit naturally the Dyson-Schwinger equations, and for the matter sector in QCD it is the quark gap equation.
The quark gap equation is schematically shown in \Cref{fig:quark-DSE}, which is a self-consistent equation for the quark propagator $G_q$, gluon propagator $G_A$ and the quark-gluon interaction vertex $\Gamma$. In the momentum space, the gap equation reads up to renormalisation constants as follows:
\begin{align}
& G_q^{-1}(p) = G_{q,0}^{-1}(p) + \Sigma(p), \qquad \textrm{with} \quad p = (\omega_p,\spatial{p}), \label{eq:quark-gapeq}  \\
& \Sigma(p) = \frac{4}{3} \, g_s \sum_{\omega_{q}} \int \frac{d^3 \spatial{q}}{(2\pi)^4} \, \gamma_\mu G_q(q)\Gamma_{\nu}(q,p)[G_A]_{\mu\nu}(q-p)\,, \label{eq:quark-gapeq-self}
\end{align}
with $g_s$ the strong coupling constant, $G_{q,0}$ the classical, bare quark propagator, and $\Sigma$ the quark self energy evaluated from a thermal sum of the 1-loop diagram in \Cref{fig:quark-DSE} over the Matsubara frequencies $\omega_{q}$ and the spatial momentum $\spatial{q}$. 
The general form for the non-perturbative quark propagator solution is as follows:
\begin{align}\label{eq:quark-prop}
 G_q^{-1}(p) &= \imag  \gamma_0  (\omega_{p} + \imag \mu_q) Z_{q}^{E}(p) \nonumber\\ 
 &+ \imag  \boldsymbol{\gamma} \cdot \boldsymbol{p} Z_{q}^{M}(p) + Z_{q}^{E}(p) M_{q}(p),
\end{align}
with $p = (\omega_p,\spatial{p})$ the quark momentum, $\omega_p$ the Matsubara frequency, $Z_q^{E,M}$ the dressed wave functions  and $M_q$ the mass function. 
Our key observation is that during the first-order phase transition, the gap equation shown in \Cref{fig:quark-DSE} also allows for an intermediate phase ($I$) between the chiral symmetry breaking (Nambu, $N$) phase and the chiral symmetric (Wigner, $W$) phase. 
This new $I$ solution branch for the quark propagator is found with an improved iteration procedure for solving the gap equation numerically, which is inspired by the homotopy method introduced in Refs.~\cite{Zheng:2023tbv} and~\cite{Wang:2012me}. The technical details of this procedure are given in \Cref{app:homo-init-iter}.

For the gap equation, we resort to the Dyson-Schwinger equations~(DSE) approach, with one of the current best truncation schemes given in Refs.~\cite{Lu:2025cls} for the gluon propagator and quark-gluon vertex in \Eq{eq:quark-gapeq-self} at finite $T$ and $\mu_{B}^{}$. Specifically, the truncation satisfies the Slavnov-Taylor identities (STIs) and the  renormalization condition in the quark-gluon vertex, see also the details in Refs.~\cite{Gao:2020qsj}. 
This approach offers quantitative precision on the chiral crossover line as well as thermodynamic quantities at small chemical potential, meeting with the benchmark results from lattice QCD and the functional renormalisation group approach. Here we  apply the scheme to further explore the higher chemical potential region. Specifically, we   focus on the $N_f=2+1$ flavor case in this work, where the $u$, $d$ and $s$ quarks for \Eq{eq:quark-prop} have the same chemical potential:
\begin{align}
\mu_u = \mu_d = \mu_s = \frac{\mu_{B}^{}}{3}.
\end{align}

\begin{figure}[tbp]
  \centering
  \includegraphics[width=0.471\columnwidth]{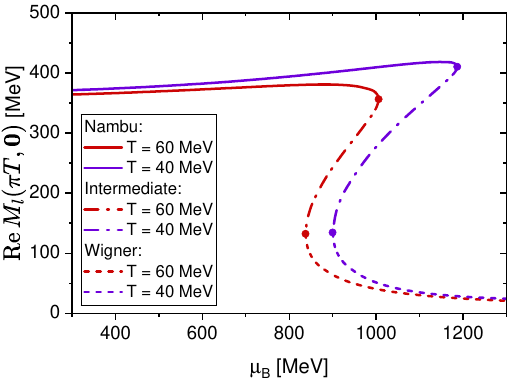}\hspace{+2ex}
  \includegraphics[width=0.48\columnwidth]{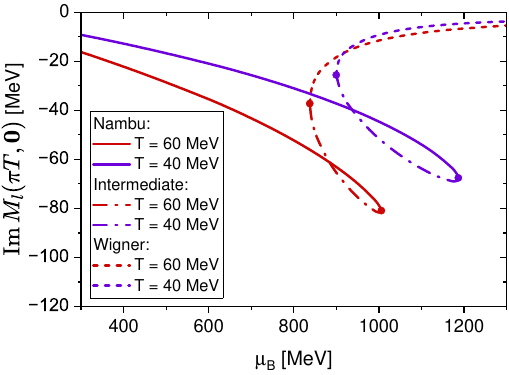}
  \caption{Light quark mass function $M_l$ ($l=u,d$) at momentum $(\pi T,\spatial{0})$: the real part (left) and the imaginary part (right), for $I$, $N$ and $W$ phases as a function of chemical potential $\mu_{B}^{}$ at temperature $T=40$ and 60\,MeV. The colored dots mark the boundaries of $I$ phase where it merges with $N$ or $W$ phase.} \label{fig:mass-branches-T}
\end{figure}

\begin{figure}[htbp]
  \centering
  \includegraphics[width=0.48\columnwidth]{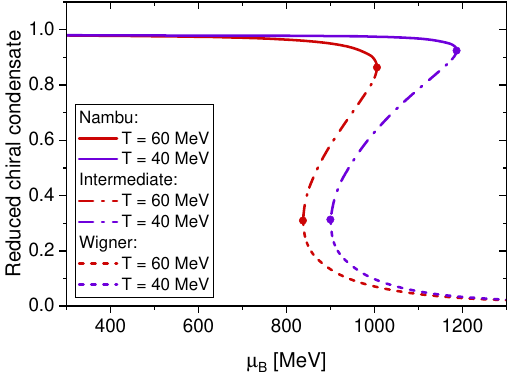}\hspace{+2ex}
  \includegraphics[width=0.48\columnwidth]{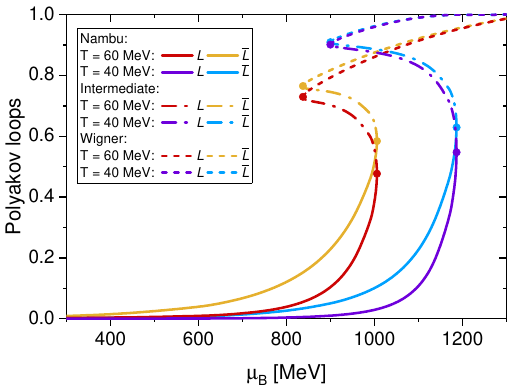}
  \caption{Left: reduced chiral condensate $\Delta_{l,s}$ for $I$, $N$ and $W$ phases as a function of baryon chemical potential $\mu_{B}^{}$ at temperature $T=40$ and 60\,MeV, which is in match with \Cref{fig:mass-branches-T}. The results are normalised by the value of $\Delta_{l,s}$ in the vacuum, i.e. at $(T,\mu_{B}^{})=(0,0)$. Right: Polyakov loops $L(\varphi_3,\varphi_8)$ and $\bar{L}(\varphi_3,\varphi_8)$, evaluated from the gluonic background field components $\varphi_3$ and $\varphi_8$ for $I$, $N$ and $W$ phases as functions of baryon chemical potential $\mu_{B}^{}$ at temperature $T=40$ and 60\,MeV, which are in match with the left panel.} \label{fig:order-params-branches-T}
\end{figure}

Within the coexistence region of $N$ and $W$ phase in first-order phase transition, there is one unique solution $G_{q}^{I}$ found other than $G_{q}^{N}$ and $G_{q}^{W}$ for the gap equation, whose respective real part of the mass function $M_q$ obeys the order:
\begin{align}
 \Re M_q^{N} > \Re M_q^{I} > \Re M_q^{W} > 0.
\end{align}
Note that we shall only consider the solutions with a positive real part of the mass function in this work. 
We specifically show the case at $T=60$ and 40\,MeV, for the light quark mass function $M_l$ ($l=u,d$) evaluated at $p = (\pi T,\spatial{0})$ as a function of baryon chemical potential $\mu_{B}^{}$, in \Cref{fig:mass-branches-T}.
It is also observed that the propagator solution $G_q^{I}$ merges with $G_q^{N/W}$ on the Nambu/Wigner phase boundary, which is in match with in the overlap of quark mass function $M_q$ as illustrated in \Cref{fig:mass-branches-T}.
Correspondingly, the chiral condensate, which is the order parameter for chiral phase transition, shows a similar behaviour. 
Here we illustrate this with the reduced chiral condensate $\Delta_{l,s}$, which is a regularised condensate defined as:
\begin{align}\label{eq:redcued-cond}
 \Delta_{l,s} &= \Delta_{l} - \frac{m_l}{m_s} \Delta_s\,, \\
 \Delta_{q} &= -T \sum_{\omega_p} \int \frac{\mathrm{d}^{3}\spatial{p}}{(2\pi)^{3}} \, \tr\, [ G_{q}(p) ] \,,
\end{align}
with $m_l$ and $m_s$ the current quark masses for light quarks $l=u,d$ and strange quark, 
see e.g. Refs.~\cite{Lu:2023mkn,Gunkel:2021oya,HotQCD:2018pds,Borsanyi:2010bp} for further details. 
Since the strange quark mass (function) is found to be quite close to its vacuum counterpart within the $\mu_{B}^{}$ region shown in \Cref{fig:mass-branches-T}, the reduced condensate offers a probe on the chiral symmetry breaking for the light quarks.
We show the $\mu_{B}^{}$ dependence of $\Delta_{l,s}$ in the left panel of \Cref{fig:order-params-branches-T}, where the three branches correspond to the propagators for the $N$, $I$ and $W$ phase that match with those at the two temperatures in \Cref{fig:mass-branches-T}. 
Such a phase structure typically demonstrates a scenario of spinodal decomposition.

The spinodal region of the chiral phase structure has further impacts on the confinement-deconfinement aspect. To see this, we evaluate the gluonic background field which is a temporal gauge field with two Cartan components $\varphi_3$ and $\varphi_8$ in color $SU(3)$: 
\begin{align}
  \varphi_{3,8} = \frac{2 \pi T}{g_s} A_0^{3,8}. \label{eq:A0}
\end{align}
Here $A_0$ denotes the temporal gauge field and $g_s$ is the strong coupling constant. 
According to a recent DSE studies~\cite{Lu:2025cls}, the gluonic background signatures the confining dynamics both in the phase structure and in thermodynamic observables. 
The latter aspect shall be further investigated in \Cref{sec:thermo}. 
As for the phase structure, the traced Polyakov loops defined by the gluonic background:
\begin{align}
  & {L}(\varphi_3,{\varphi}_8) = \frac{1}{3}\left[e^{- \frac{2 \pi}{\sqrt{3}}\imag {\varphi}_8} + 2 \, e^{\frac{\pi}{\sqrt{3}} \imag {\varphi}_8} \cos{\pi\varphi_3}\right]\,, \\
  & \bar{L}(\varphi_3,{\varphi}_8) = {L}(\varphi_3,-{\varphi}_8)\,,
\end{align}
manifests the center symmetry aspect which provides a proxy for the confinement-deconfinement phase transition. 
The gluonic fields $\varphi_3$ and $\varphi_8$ as functions of $T$ and $\mu_{B}^{}$ are determined by their equations of motions, which correspond to the stationary point in the Polyakov loop potential. The computational details can be found in the Appendix A of Refs.~\cite{Lu:2025cls}. 
The respective Polyakov loops $L$ and $\bar{L}$ for $N$, $I$ and $W$ phase are shown in the right panel of \Cref{fig:order-params-branches-T} , which are in match with \Cref{fig:mass-branches-T}.
Similar to the chiral condensate, the Polyakov loops are also found with a spinodal-type transition in the coexistence region where $I$ phase is connected to $N$ phase and $W$ phase on each phase boundary. 
In particular, the phase boundary in the Polyakov loops coincides with the ones for the chiral condensate $I$, which generalises our previous finding in Refs.~\cite{Lu:2025cls} that the chiral and confinement phase transition are closely related not only in the vicinity of CEP but also beyond that in the region of first-order phase transition.

\begin{figure}[tbp]
  \centering
  \includegraphics[width=0.7\columnwidth]{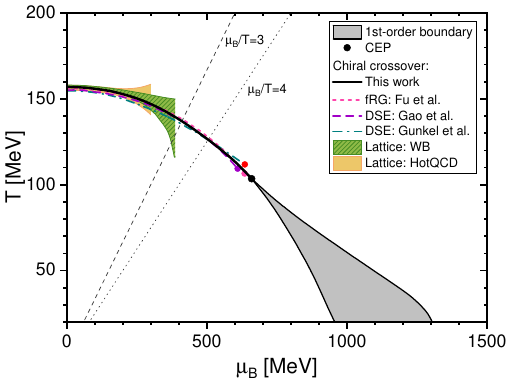}
  \caption{QCD phase diagram in the first-order transition region. The phase boundaries are defined by the existence boundaries of $I$ phase, which is shown as the boundaries of the grey area. The state-of-the-art results on the chiral crossover line their estimated critical end point are also put in, including the lattice QCD \cite{Borsanyi:2020fev,HotQCD:2018pds} and the functional QCD results (DSE:~\cite{Gao:2020fbl,Gunkel:2021oya}, fRG:~\cite{Fu:2019hdw}), together with the estimates on the CEP location (colored dots).} \label{fig:phase-diagram}
\end{figure}

In short, we arrive at a QCD phase diagram shown in \Cref{fig:phase-diagram}, where the phase boundaries of first-order transitions are defined by the boundaries for $I$ phase at different temperatures, which is shown as the grey area in the figure. 
It is found that the $\mu_B$ width between the two phase boundaries gets increased when temperature $T$ decreases. In turn, the spinodal region gets larger at lower $T$, which also indicates that the supercooling and overheating effects shall get enhanced. 
Together, the state-of-the-art results of the chiral crossover line calculated from lattice QCD~\cite{Borsanyi:2020fev,HotQCD:2018pds} and the functional QCD approaches (DSE:~\cite{Gao:2020fbl,Gunkel:2021oya}, functional renormalisation group - fRG:~\cite{Fu:2019hdw}) are also put in. 
We also mark out the state-of-the-art estimates for the CEP location as the colored dots in \Cref{fig:phase-diagram}, which are given by direct calculations in the functional approaches from the literatures listed above.  
In turn, the present work is within these up-to-date computational frameworks of functional QCD, which does not provide a new estimate on the location of CEP. 
Notice that the phase boundary with a lower $\mu_B$ are getting close to the region for the nuclear liquid-gas phase transition. 
Within the present DSE truncation scheme, the liquid-gas phase transition is not incorporated as the emergent hadronic degrees of freedom at high density have not been explicitly included. The investigation on the latter aspect is still preliminary in functional approaches, see e.g.~\cite{Gao:2025kzk,Fukushima:2023wnl}, and a fully systematic evaluation on that in QCD is planned in future works. 
As we shall see in the next Section, this shall leave an impact on the predictions of thermodynamic properties for the cold and dense strong interaction matter. 
Typically, the phase diagram in the temperature--baryon-density ($n_B$) plane is qualitatively different at low temperature between the case that one includes or excludes the emergence of liquid-gas phase transition.
A phenomenological incorporation of the liquid-gas phase transition will be put forward in the next section, where its impact is also discussed. 
We also note that we have left out the possible new phases in QCD at high density such as the spatial modulations (moats or inhomogeneity), color superconductivity and so on, which indicate the possibility of novel condensates and emergent degrees of freedom above the onset regime of CEP. 
The incorporation on these rich phase structures is beyond the scope of this work. 
Here we would like to simply focus on the observed spinodal decomposition in the chiral and confinement-deconfinement phase structure of \textit{homogeneous} strong interaction matter, and then discuss its impact on the thermodynamic observables, which are presented in the following Section. 
Besides, by further incorporating a phenomenological   description of the bulk inhomogeneity  between the two phases during the first order phase transition, the present knowledge allows us to investigate the formation of bubbles in dense strong interaction matter.

\section{Equation of state and thermodynamic observables at low temperature and high density}\label{sec:thermo}

We continue to apply the knowledge of spinodal decomposition to the study of thermodynamic observable at low temperature and high density. 
Here, a straightforward observable is the net-quark number density, which is directly accessible from the normalised quark propagator $\bar{G}_q$ and the gluonic background field $\varphi$ in \Eq{eq:A0}~\cite{Lu:2025cls}:
\begin{align}
n_{q}(T,\mu_q) &= - T \sum_{\omega_p} \int \frac{\mathrm{d}^{3}\spatial{p}}{(2\pi)^{3}} \tr\,\bigl[ \gamma_0\, \bar{G}_{q}( p^{\varphi} )  \bigr]\,, \label{eq:net-density} \\
p^{\varphi} &= ( \omega_p + 2\pi T \varphi, \spatial{p} )\,. \label{eq:mom-with-gluonic-field}
\end{align}
In terms of the Cartan field components, the gluonic field in \Eq{eq:mom-with-gluonic-field} takes the eigenvalues of the fundamental representation:
\begin{align}
  \varphi \in \bigl( \frac{\varphi_3}{2} + \frac{\varphi_8}{2\sqrt{3}}, -\frac{\varphi_3}{2} + \frac{\varphi_8}{2\sqrt{3}}, -\frac{\varphi_8}{\sqrt{3}} \bigr).
\end{align}
when taking the color trace in \Eq{eq:net-density}. 
The normalisation of the quark propagator is set by $\bar{G}_q = Z_q^{E}\, G_{q}$ with the thermal wave function dressing $Z_q^{E}$ defined in \Eq{eq:quark-prop}. 
The net-quark number further relates to the net-baryon number as a conserved charge, whose density follows $n_{B}^{} = (n_u+n_d+n_s)/3$.
In the crossover region, \Eq{eq:net-density} already has some decent applications on studying the QCD equation of state, baryon number fluctuations and so on, see e.g. Refs.~\cite{Lu:2025cls,Lu:2023mkn,Isserstedt:2019pgx} via the DSE approach. 
In this work, we further extend the calculation of \Eq{eq:net-density} to the first-order transition region, with both the chiral dynamics and the confining dynamics~(\Cref{fig:order-params-branches-T}) incorporated self-consistently. 

\begin{figure}
  \centering
  \includegraphics[width=0.48\columnwidth]{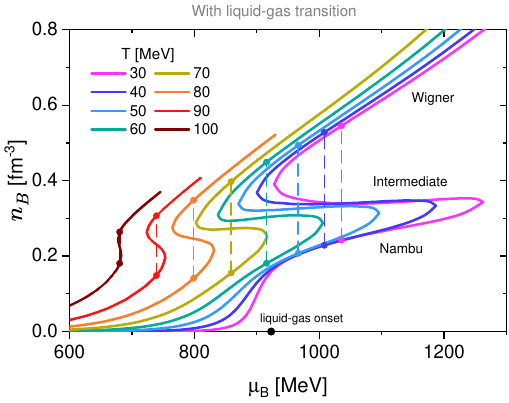}\hspace{2ex}
  \includegraphics[width=0.475\columnwidth]{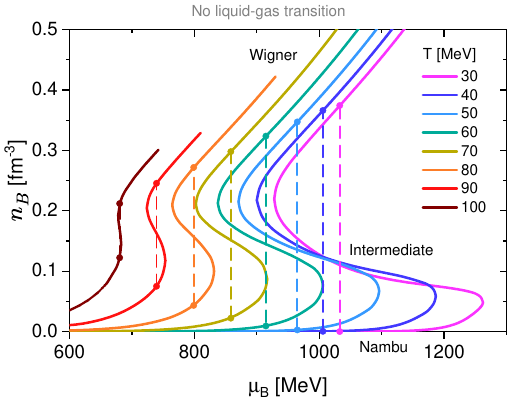}
  \caption{Net-baryon number density $n_B$ as a function of baryon chemical potential at different temperatures, with a multi-phase structure during first-order phase transition, for temperatures $T=30$ to $100\,$MeV in a 10\,MeV step. The left and the right panel show a comparison where the liquid gas phase transition occurs or not, respectively. The Maxwell construction is indicated by the dashed vertical lines, whose end points (solid dots) indicate the boundary condition for the inhomogeneous nuclear density distribution. }\label{fig:net-density}
\end{figure}

Similar to the behaviour shown in the order parameters, we found that the net-baryon number density for $N$, $W$ and $I$ phase at given temperature shows a multi-phase structure during the first-order phase transition. 
The results are summarised in the right panel of \Cref{fig:net-density}, where we demonstrate the chemical potential dependence of $n_{B}^{}$ at fixed temperature for all possible phases. 
Note that the results can also be regarded as the isothermal trajectories $T(n_B,\mu_B) = \textrm{constant}$.
With no liquid-gas transition occurs in this scenario, the resulted $n_B$ from the net-quark number densities through the propagator shows a trend of being vanishing on the lower boundary of the coexistence region. 
At zero temperature, this behaviour is well understood as the Silver-Blaze property~\cite{Cohen:2003kd,Gunkel:2019xnh,Gunkel:2020wcl}: if there is no explicit singularity located in the complex $\tilde{p}_0 = p_0 + \imag \mu$ plane for $\mu \in [0,\mu_q]$, where $p_0$ is the temporal component of the Euclidean quark momentum, then the quark propagator shall follow the analytic continuation relation:
\begin{align}
G_q(p_0,\spatial{p};\mu_q) = G_q((p_0+\imag\mu_q)^2+\spatial{p}^2) = G_q(p_0+\imag \mu_q,\spatial{p};0). 
\end{align}
Accordingly, all thermodynamic observables remain the same as their vacuum expectation values, typically the net-quark number density shall remain to be vanishing.
In reality, the dense strong interaction matter should first undergo a liquid-gas phase transition, typically with a critical chemical potential $\mu_B = 923\,$MeV at zero temperature, which according to our computation is below the Nambu phase boundary. This corresponds to a pole in the nucleon propagator and hence, the Silver Blaze property shall not preserved when $\mu_B$ goes beyond the critical chemical potential of liquid-gas transition, see e.g.~\cite{Gao:2025kzk} for a qualitative evaluation on this through the quark self energy. 

As mentioned in the previous section, in this work we are satisfied in providing a phenomenological description on the liquid-gas phase transition for studying its potential overlap with the QCD phase transition. 
To achieve this, we combine the nuclear matter equation of state with a liquid-gas transition together with the above mentioned DSE thermodynamic function for quarks where there is no liquid-gas transition related poles and singularities. 
For the hadronic part, we refer to the Walecka model as a minimal construction on the nuclear liquid-gas phase transition, where the nucleons $N$, scalar mesons $\sigma$ and vector mesons $\omega_0$ are taken into account. Under the mean-field approximation on the mesonic fields, its (Euclidean) Lagrangian reads:
\begin{align}
 \mathcal{L}_{\textrm{hadron}} = \bar{N} [ - \gamma \cdot \partial - \gamma_0 (\mu_B - g_{\omega} \omega_0) + (M_N - g_{\sigma} \sigma) ] N - \frac{1}{2} m_{\sigma} \sigma^2 + \frac{1}{2} m_{\omega} \omega_0^2.  \label{eq:Walecka-MF}
\end{align}
At finite temperature and finite chemical potential, the mesonic fields are determined by the following gap equations self-consistently: 
\begin{align}
 \sigma &= \frac{g_{\sigma}}{m_{\sigma}^{2}} \langle \bar{N}{N} \rangle = \frac{4 g_{\sigma}}{m_{\sigma}^{2}} \int \frac{\mathrm{d}^3 \spatial{p}}{(2\pi)^3} \, \frac{M^{*}}{E^{*}(\spatial{p})} f(E^{*}(\spatial{p}) - \mu_B^{*}), \\
 \omega_0 &= \frac{g_{\omega}}{m_{\omega}^{2}} \langle \bar{N} \gamma_0 {N} \rangle = \frac{4 g_{\omega}}{m_{\omega}^{2}} \int \frac{\mathrm{d}^3 \spatial{p}}{(2\pi)^3} \, f(E^{*}(\spatial{p}) - \mu_B^{*}), 
\end{align}
where $M_{N}^{*} = M_{N} - g_{\sigma} \sigma$ and $\mu_{B}^{*} = \mu_B - g_{\omega} \omega_0$ are the effective nucleon mass and chemical potential, $E^{*}(\spatial{p}) = \sqrt{\spatial{p}^2 + (M_{N}^{*})^2}$ is the energy dispersion, and $f$ represents the Fermi-Dirac distribution function. 
Each of the mass parameter matches the vacuum hadron mass $M_N = 938\,$MeV, $m_{\sigma} = 550\,$MeV and $m_{\omega} = 783\,$MeV. 
For the Yuwaka couplings, we follow the setup in Ref.~\cite{Walecka:1974qa} to take  $g_{\sigma} = 16.3 \, m_{\sigma} / M_N$ and $g_{\omega} = 14.0 \, m_{\omega} / m_N$, to wit the nucleon saturation density $n_0 = 0.16\,\textrm{fm}^{-3}$ and a binding energy of 15.8\,MeV at zero temperature. 
Thermodynamic quantities are further available once the mesonic fields are determined. For example, the nucleon density reads $\langle \bar{N} \gamma_0 {N} \rangle =  \omega_0 \, m_{\omega}^2/g_{\omega}$ which will be specified as the hadronic contribution of the baryon density. Also, the pressure $P$ and energy density $\epsilon$ of nucleon matter are given by:
\begin{align}
  P &= - \frac{1}{2} m_{\sigma} \sigma^2 + \frac{1}{2} m_{\omega} \omega_0^2 + 4 \int \frac{\mathrm{d}^3 \spatial{p}}{(2\pi)^3} \, E^{*}(\spatial{p}) f(E^{*}(\spatial{p}) - \mu_B^{*}), \\
  \epsilon &= \frac{1}{2} m_{\sigma} \sigma^2 + \frac{1}{2} m_{\omega} \omega_0^2 + \frac{4}{3} \int \frac{\mathrm{d}^3 \spatial{p}}{(2\pi)^3} \, \frac{\spatial{p}^2}{E^{*}(\spatial{p})} f(E^{*}(\spatial{p}) - \mu_B^{*}).
\end{align}

The hadronic and quark contributions of the equation of state are then mixed under the excluded volume mechanism, which follows similarly as the framework introduced in Refs.~\cite{Steinheimer:2010ib,Steinheimer:2011ea}. This results in the modified baryon chemical potentials for the hadronic contribution with effective repulsion:
\begin{align}
\tilde{\mu}_{B} = \mu_{B} - v P_{\textrm{tot.}},
\end{align}
with $P_{\textrm{tot.}}$ the total pressure of hadrons and quarks. Typically the excluded volume size is chosen as $v = 1\,\textrm{fm}^{3}$ in Ref.~\cite{Steinheimer:2010ib} which is also adopted here. Meanwhile, quarks have no repulsive corrections in their chemical potentials, namely their excluded volume factors are zero.
All thermodynamic quantities should then be understood as those at the modified chemical potential $\tilde{\mu}$. In addition, the actual densities of thermodynamics shall receive a volume correction factor $f$: 
\begin{align}\label{eq:excvolume}
f = \frac{V^{\prime}}{V} = (1+ v n_{B} )^{-1},
\end{align}
such as the number densities $\tilde{n}_B = f \, n_B$, $\tilde{n}_q = f \, n_q$, the total energy density $\tilde{\epsilon}_{\textrm{tot.}} = f \, \epsilon_{\textrm{tot.}} = f(\epsilon_{\textrm{hadron}} + \epsilon_q)$ and similarly for the entropy density and so on. In particular, the pressure follows the thermodynamic relations $\tilde{P} = T \tilde{s}_{\textrm{tot.}} + \sum_{i} \tilde{\mu}_{B,i}^{} \tilde{n}_{B,i}^{} - \tilde{\epsilon}_{\textrm{tot.}}$ with $i$ specifying all particle components. 
For convenience, we shall omit the ``\,$\tilde{~}$\,'' and ``tot.'' notations for specifying the actual total thermodynamic quantities obtained from the mixing approach. To distinguish from the thermodynamic quantities with only the quark sector included, we would refer to whether the case has incorporated the liquid-gas phase transition or not. 

Note that since we have adopted the mean-field approximation for the meson fields in the Walecka model, i.e. the mesonic fields are integrated out to constitute the effective mass as well as the chemical potential for the nucleons as shown in the mean-field Lagrangian in \Eq{eq:Walecka-MF}, hence we regard that there will be no volume correction from meson densities in \Eq{eq:excvolume}. 
To emphasise, our motivation of such a mixing construction is slightly different from what is originally put forward in Refs.~\cite{Steinheimer:2010ib,Steinheimer:2011ea}, 
since here the main purpose is to incorporate a qualitative reliable description on the Nambu phase with a liquid-gas phase transition, which is done within approximation instead of capturing that upon the level of dynamical quarks and gluons. 
In fact, preliminary studies show that there are indeed structural similarities between the baryonic loop diagram in the quark gap equation and the role of $\sigma$ and $\omega$ in the Walecka model, which shifts the nucleon mass and chemical potential, see more details in Ref.~\cite{Gao:2025kzk}. 
In turn, this brings us an opportunity to study the implicit influence of the nuclear liquid-gas pole on the thermodynamics of strong interaction matter at high density. 

\begin{figure}[tbp]
  \centering
  \includegraphics[width=0.65\columnwidth]{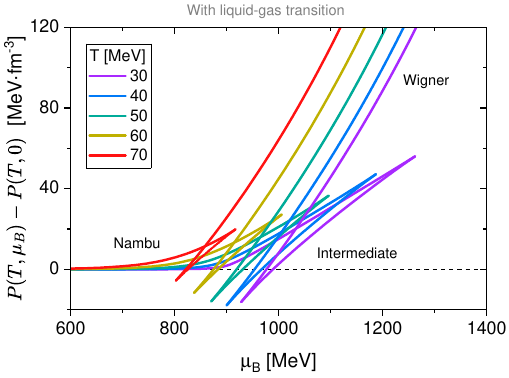}
  \caption{The pressure difference $P(T,\mu_B) - P(T,0)$ as a function of baryon chemical potential $\mu_B$ for Nambu, Wigner and Intermediate phase at $T=30,40,50,60$ and $70\,$MeV during the first-order QCD phase transition, computed via \Eq{eq:pressure}. In this case, the liquid-gas phase transition is incorporated.}\label{fig:pressure}
\end{figure}

First, we investigate the total net-baryon number density. In the left panel of \Cref{fig:net-density} we show the total $n_{B} = f(n_B^{\textrm{hadron}} + n_B^{\textrm{quark}})$ followed by the volume correction factor in \Eq{eq:excvolume}, which is computed for all different phases of chiral symmetry, namely Nambu, Wigner and Intermediate phase. 
Due to the baryon number excitation from the liquid-gas transition, the Nambu phase acquires a density higher than the saturation density above $\mu_B = 923\,$MeV, which qualitatively different from the case with no liquid-gas transition where the density is greatly suppressed around that chemical potential. 
The trend of the isothermal $n_B$ trajectories dictates that such influence from the liquid-gas pole persists even for temperatures much higher than the critical point of liquid-gas transition (around 15\,MeV), up to around $60\,$MeV. While beyond that temperature or at a chemical potential much smaller than 923\,MeV, the influence of the liquid-gas transition on $n_B$ becomes less relevant.
As a consequence, the density range of the intermediate phase is narrowed, and the area of the spinodal region shrinks. 
We also notice that for the mixed total baryon number density in the Intermediate phase, the susceptibility $\partial n_B /\partial \mu_B$ switches from negative to positive and then to negative along the isotherm. This would lead to a divergent speed of sound at the points where the susceptibility flips sign, which is unphysical and leaves as a caveat on the present mixing approach. This also suggests that the nuclear liquid gas transition may also have synergy with the spinodal decomposition of chiral phase transition, which leads to stronger fluctuations in the first-order phase transition. 
Nevertheless, the realistic path for a first-order chiral phase transition should follow the coexistence line where the Nambu phase and the Wigner phase share an equal pressure and chemical potential, hence we regard that the peculiar thermodynamics in the Intermediate phase only amounts to physics inside the spinodal region, while they shall be bypassed through the phase coexistence. 

\begin{figure}
  \centering
  \includegraphics[width=0.65\columnwidth]{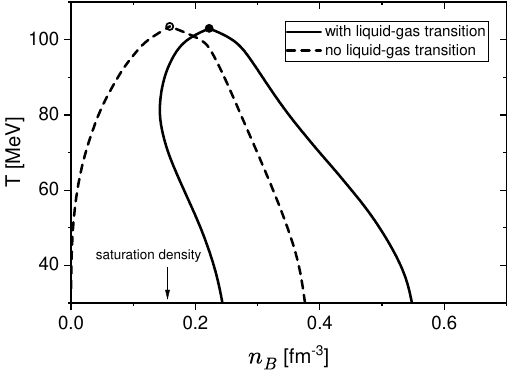}
  \caption{Phase diagram for the chiral phase transition below the temperature of CEP (dots in the plot) in the $(n_B,T)$ plane. The net-baryon density $n_B$ of the Nambu phase and the Wigner phase along the coexistence line are shown as the two curves respectively, which are determined from the Maxwell construction~\Eq{eq:MC}. The case with a liquid-gas phase transition (solid) and the case without that (dashed) are compared together with the nuclear saturation density $n_0 = 0.16\,\textrm{fm}^{-3}$ (arrow position).}\label{fig:phase-diagram-nB}
\end{figure}

Given a connected isothermal trajectory in the $(\mu_B,n_B)$ plane, the coexistent chemical potential follows from the Maxwell construction (MC) as: 
\begin{align}\label{eq:MC}
\int_{\tau_N}^{\tau_W} n_B(\tau) \frac{\mathrm{d} \mu_{B}(\tau)}{\mathrm{d} \tau} \mathrm{d} \tau = P_{W} - P_{N} = 0,
\end{align}
where $\tau$ is the control parameter on the isotherm, and the end points of the integral satisfy the boundary condition $\mu_{B}(\tau_{N}) = \mu_{B}(\tau_{W}) = \mu_{B}^{\textrm{MC}}$ at the coexistent chemical potential. 
Note that \Eq{eq:MC} is a generalized representation on the Maxwell construction as put forward in Ref.~\cite{Mei:2025ein}, and the most commonly used representation is readily incorporated if one takes $\tau = \mu_B$ and perform the integral in \Eq{eq:MC} on the path along the isotherm. 
In fact, this parametrization also allows us to compute the pressure (difference) at a given $T$ for all phases within the spinodal region:
\begin{align}\label{eq:pressure}
  P(\mu_{B,2}) - P(\mu_{B,1}) = \int_{\tau_1}^{\tau_2} n_B(\tau) \frac{\mathrm{d} \mu_{B}(\tau)}{\mathrm{d} \tau} \mathrm{d} \tau, \qquad \mu_{B,i} = \mu_B(\tau_i).
\end{align}
We specifically show the result of pressure difference $P(T,\mu_B) - P(T,0)$ with a liquid-gas phase transition in \Cref{fig:pressure}. The qualitative feature of the pressure in different phases is vital for understanding the interface effect during a first-order QCD phase transition, and the respective study shall be given in the next Section.

In both cases with or without the liquid-gas transition, the Maxwell construction is applied with the obtained $\mu_{B}^{\textrm{MC}}$ positions represented by the dashed vertical lines in \Cref{fig:net-density}. 
Some crucial thermodynamic properties can then be read out. For example, one can further extract the phase diagram in the $(n_B,T)$ plane as shown in \Cref{fig:phase-diagram-nB}, with a coexistence region that represents the allowed $n_B$ range for $I$ phase at different temperatures, and its upper and lower boundaries are in match with those of the first-order phase transition region shown in \Cref{fig:phase-diagram}. Particularly, we see that the impact of a liquid-gas phase transition gives a qualitative difference on the lower boundary of the phase diagram, which in our case could result in a ``back-bending'' shape towards zero temperature in order to reach the saturation density $n_0$. Note that the possibility of such a back-bending scenario has also been addressed in Ref.~\cite{Steinheimer:2012gc}, and in our case this relates to the fact that the state-of-the-art QCD estimate on the baryon density at CEP already points to the magnitude of $n_0$. 

Moreover, we extract the speed of sound in the respective low temperature and high density region, which is highly relevant to the neutron star equation of state and also the high baryon density matter to be produced in CBM experiment. 
\begin{figure}
  \centering
  \includegraphics[width=0.48\columnwidth]{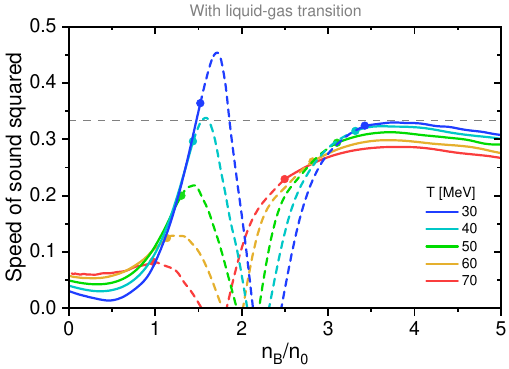}\hspace{2ex}
  \includegraphics[width=0.48\columnwidth]{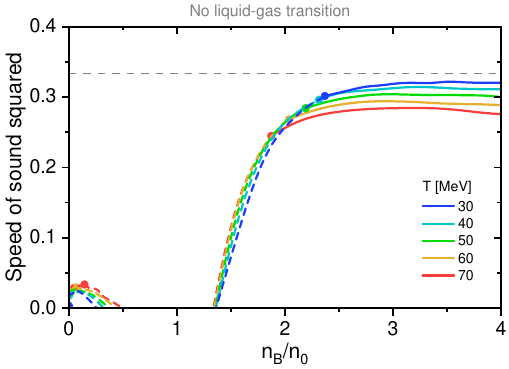}
  \caption{Isotherm speed of sound squared $(\partial P / \partial \epsilon)_{T}$ as a function of baryon number density $n_B$ for $T \in (30,70)\,$MeV. The case of a reliable hadronic phase with the liquid gas transition (left panel) and the case without that (right panel) are both investigated. The dots stands for the coexistence boundary of Nambu and Wigner phase, and the computation covers the range of the Nambu or Wigner phase away from the coexistence boundary (solid curves) and the range from the coexistence boundary to the first-order phase boundary (dashed curves), for baryon density below 5 times of the nuclear saturation density $n_0$. The conformal limit of the speed of sound squared $1/3$ is marked by the gray dashed line.}\label{fig:speed-of-sound}
\end{figure}
In \Cref{fig:speed-of-sound} we demonstrate the isothermal speed of sound squared:
\begin{align}\label{eq:speed-of-sound}
  c_T^2 = \bigl( \frac {\partial P}{\partial \epsilon} \bigr)_T = \frac{n_B}{ T \frac{\partial s}{\mu_B} + \mu_B \frac{\partial n_B}{\mu_B} }, 
\end{align}
as a function of net-baryon number density across the first-order phase transition (solid curves) in the range of $T \in (30,70)\,$MeV, where the cases with or without the liquid-gas phase transition are both investigated. 
In the case with liquid-gas phase transition, the mixing of hadronic thermal densities and quark thermal densities is considered as discussed around \Eq{eq:excvolume}. Note that the minimum shown in the low temperature case at around $n_B/n_0 < 1$ corresponds to the critical point of the liquid-gas transition, around which the equation of state is soften. 
It is clear from the comparison that the emergence of the liquid-gas transition enhances greatly the speed of sound around the onset density $n_B/n_0 \approx 1$ at low temperature, before the first-order QCD phase transition comes into play which tends to decrease the speed of sound. 
Especially, the speed of sound is strongly suppressed and becomes vanishing at the Nambu phase boundary and the Wigner phase boundary, as the dashed curves shown in \Cref{fig:phase-diagram} which covers the range from the coexistence boundary (dot) to each of the phase boundary. 
The latter suppression of the speed of sound around QCD phase boundary is also seen in the scenario where there is no liquid-gas transition, in the right panel of \Cref{fig:phase-diagram}. However, in this case the speed of sound shall never reach above the conformal limit $1/3$, and in particular in the Nambu phase the speed of sound is quite small. 
Together, our picture stands that if a successive first-order QCD phase transition occurs after the nuclear liquid-gas transition, the equation of state would first grows stiff and then becomes soft and finally goes stiff again to bypass all the phase transitions. Such a scenario is in accordance with the estimate from the Bayesian analysis on heavy-ion collisions data~\cite{Oliinychenko:2022uvy}. 
While we note that the possibility also lies there that, if the QCD phase transition is smeared to a crossover at high density due to the interference of the liquid-gas pole or other aspects, then it is possible that the equation of state keeps stiff with a speed of sound squared compatible or larger than the conformal limit, which is reflected in the estimations from neutron star observables~\cite{Tews:2018kmu,Tews:2019ioa,Altiparmak:2022bke,Fujimoto:2024cyv} or other estimations from heavy-ion collisions~\cite{OmanaKuttan:2022aml,Chatterjee:2023ecc}. Note that these estimations correspond to the speed of sound at zero temperature. 
Of course, we have to note that the present investigation is a case study that singles out two main aspects near the onset of nucleon saturation density, which are the nuclear liquid-gas phase transition onset and the interwind chiral symmetry breaking of quarks with the center symmetry breaking of gluonic backgrounds. 
As we already stated around the discussion on the phase diagram, there are other possible emergences at high density such as the color-superconducting condensate, which can also have impact on the equation of state and the speed of sound, see e.g. Refs.~\cite{Geissel:2024nmx,Geissel:2025vnp}. Hence, approximation is adopted here that we expect other emergent degrees of freedom will only play a role at a much larger baryon density.

\section{Interface effect in the spinodal decomposition region}\label{sec:interface}

It is well known that the interface would induce the supercooling and overheating phenomena during the first-order phase transition. For QCD phase transition, a possible outcome from that is the formation of interface between the coexistent Nambu and Wigner phase~\cite{Ke:2013wga,Gao:2016hks,Mei:2025ein}. 
The investigation on QCD thermodynamic quantities in the previous section allows us to gain further sights on the formation of such an inhomogeneous structure, and especially on the implicit influence of liquid-gas phase transition on that at low temperature. 

\begin{figure}[tbp]
  \centering
  \includegraphics[width=0.6\columnwidth]{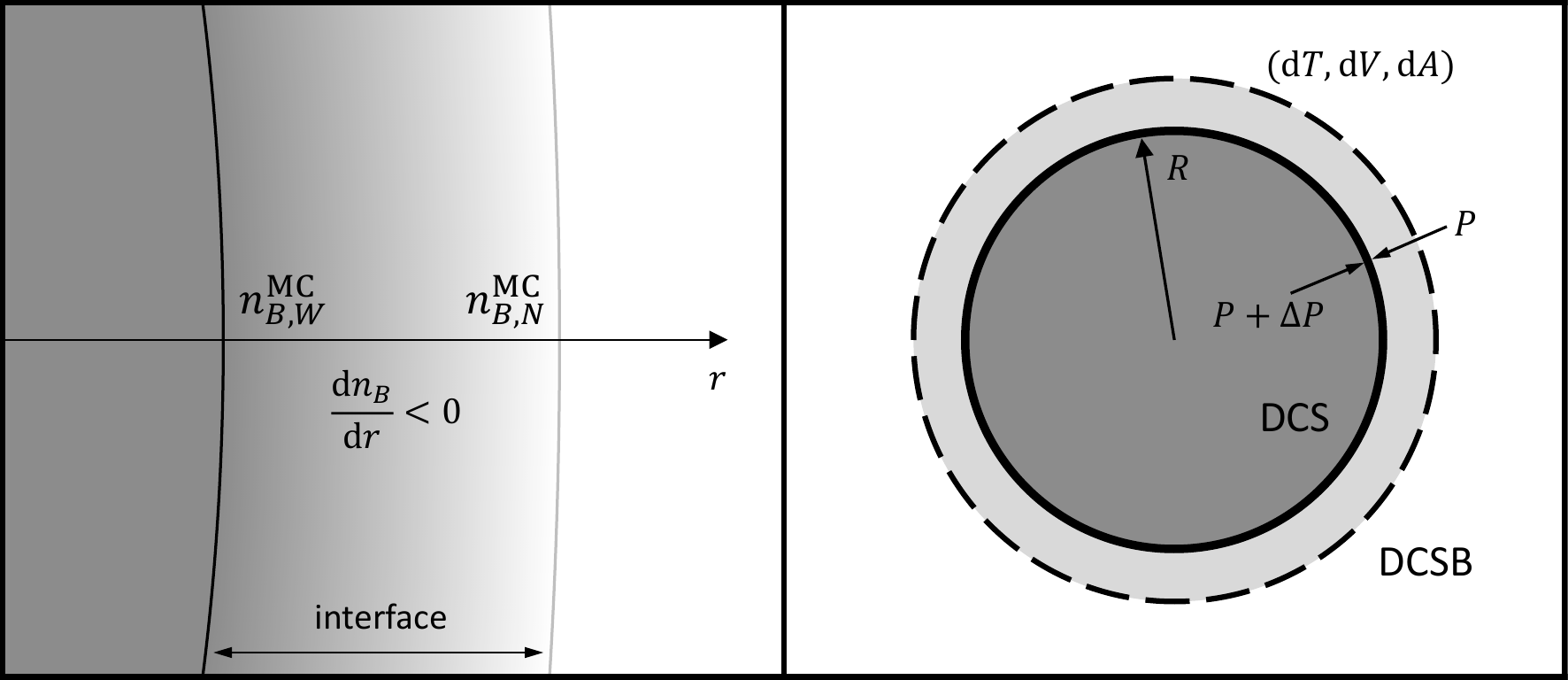}
  \caption{Left panel: sketch on the interface structure for co-existent Nambu phase and Wigner phase in dense nuclear matter. The net-baryon densities at the two boundaries are marked out as $n_{B,N}^{\textrm{MC}}$ and $n_{B,W}^{\textrm{MC}}$. We specifically show the case of a negative density gradient $\mathrm{d}n_{B}^{}/\mathrm{d}r<0$ indicated by the color, which describes a bubble with dense core. 
  Right panel: Formation a nuclear bubble with a curved interface: the solid circle represents the whole interface shown in the left panel, which separates the Nambu and the Wigner phase with their pressure difference $\Delta P$. The interface expands or shrinks due to the change of temperature $T$, volume $V$ or surface area $A$, which is illustrated by the light-gray area and the dashed circle.}\label{fig:interface}
\end{figure}

A schematic picture of the interface formation is shown in the left panel of \Cref{fig:interface}. On each side of the interface, we have the homogenous Nambu phase and Wigner phase, which share the same chemical potential and pressure under the thermodynamic equilibrium. This follows exactly the coexistence condition described by the Maxwell construction as mentioned in the previous section, hence the baryon density at the interface boundary correspond to:
\begin{align}
  n_{B,N}^{\textrm{MC}} &= n_{B,N}^{}(T,\mu_{B}^{\textrm{MC}}), \nonumber \\
  n_{B,W}^{\textrm{MC}} &= n_{B,W}^{}(T,\mu_{B}^{\textrm{MC}}), \label{eq:density-MC}
\end{align}
respectively. While inside the interface, the baryon density changes gradually from one to another together with the chemical potential, with respect to the spatial coordinate $\spatial{r}$. 
In \Cref{fig:interface} we specifically show the case for a negative density gradient $\mathrm{d}n_{B}^{}/\mathrm{d}r$ with respect to the interface normal direction $r$. 
The inhomogeneous distribution of the density $n_{B}^{}(\spatial{r})$ shall minimise the total free energy of the interface, which can be solved by considering the stationary condition:
\begin{align}
  \int \mathrm{d}^3 \spatial{r} \, \delta f(\spatial{r}) = 0, \label{eq:energy-stationary}
\end{align}
for an arbitrary variation on the density distribution $\delta n_{B}^{}(\spatial{r})$, with $f(\spatial{r})$ the corresponding free energy density. 
We explicitly adopt the phenomenological model developed in Ref.~\cite{Randrup:2009gp} for the total energy density, which takes into account the inhomogeneous effect: 
\begin{align}
  & f(\spatial{r}) = \frac{1}{2} C (\nabla_{\spatial{r}} n_{B}^{})^2 + f_{\textrm{bulk}}(\spatial{r}). \label{eq:free-energy-dist}
\end{align}
The bulk energy density satisfies:
\begin{align}
\delta f_{\textrm{bulk}}(\spatial{r}) = \mu_{B}^{}(n_{B}) \, \delta n_{B}^{}(\spatial{r}),
\end{align}
for the variation on the density distribution. 
The chemical potential $\mu_{B}^{}$ as a function of the net-density $n_{B}^{}$ shall take input from the results under a homogeneous configuration, which has been given in~\Cref{fig:net-density}. 
The quadratic gradient term in \Eq{eq:free-energy-dist} describes the inhomogeneous contribution to the energy density, with a constant $C$ specified by the net-baryon density $n_{B,c}$ and energy density $\epsilon_c$ at the CEP, together with the thickness of the interface $a$:
\begin{align}
  C=a^2\frac{\epsilon_c}{n_{B,c}^2}.
\end{align}
According to the calculation in Ref.~\cite{Lu:2025cls}, we have $n_{B,c} = 0.158\,\textrm{fm}^{-3}$ and $\epsilon_c = 228\,\textrm{MeV}\cdot\textrm{fm}^{-3}$. 
For the thickness parameter $a$, we follow Refs.~\cite{Ke:2013wga,Gao:2016hks} to take $0.33\,\textrm{fm}$. 

\Eq{eq:energy-stationary} is then solved under the total net-baryon number conservation:
\begin{align}
  \delta N_{B}^{} = \int \mathrm{d}^3 \spatial{r} \, \delta n_{B}^{}(\spatial{r}) = 0,
\end{align}
and the solution should obey the following Poisson equation:
\begin{align}\label{eq:inhomo-eom}
  C \nabla_{\spatial{r}}^2 n_{B}^{} = \mu_{B}^{}(n_{B}^{}) - \mu_{B}^{\textrm{MC}},
\end{align}
with $\mu_{B}^{\textrm{MC}}$ introduced as the Lagrange multiplier from the net-baryon number conservation. 
In this work, we are satisfied in providing the planar interface solutions to \Eq{eq:inhomo-eom}, which are explicitly~\cite{Ke:2013wga}:
\begin{align}\label{eq:planar-interface}
\frac{\mathrm{d}n_{B}^{}}{\mathrm{d}r} = \pm \sqrt{\frac{2\Delta f}{C}},
\end{align}
with $\Delta f$ the free energy budget in the spiondal region defined by a (path) integral on the baryon density:
\begin{align}\label{eq:free-energy-spinodal}
  \Delta f(n_{B}^{}) = \int_{n_{B,N}^{\textrm{MC}}}^{n_{B}^{}} [\mu_{B}^{}(n) - \mu_{B}^{\textrm{MC}}] \, \mathrm{d}n, \quad  n_{B,N}^{\textrm{MC}} \leq n_{B}^{} \leq n_{B,W}^{\textrm{MC}}.
\end{align}
The $\pm$ sign in \Eq{eq:planar-interface} indicates the direction of the interface with respect to the one displayed in \Cref{fig:net-density}. 
The density distribution in \Eq{eq:planar-interface} further allows us to calculate the interface tension $\sigma$, which is defined as the free energy deficit of \Eq{eq:free-energy-dist} per unit area on the interface.  
This can be calculated by integrating out the spatial coordinate $r$ over the free energy density budget across the interface, as~\cite{Randrup:2009gp}:
\begin{align}\label{eq:interface-tension}
  \sigma &= \int_{-\infty}^{\infty} \big[ \Delta f + \frac{C}{2} \bigl( \frac{\mathrm{d}n_{B}^{}}{\mathrm{d}r} \bigr)^2 \bigr] \, \mathrm{d}r \nonumber \\
  &= \int_{n_{B,N}^{\textrm{MC}}}^{n_{B,W}^{\textrm{MC}}} C \left(\frac{\mathrm{d}n_{B}^{}}{\mathrm{d}r}\right)^2 \frac{\mathrm{d}r}{\mathrm{d}n_{B}^{}} \mathrm{d}n_{B}^{}  \nonumber \\
  &= \int_{n_{B,N}^{\textrm{MC}}}^{n_{B,W}^{\textrm{MC}}} \sqrt{2 C \, \Delta f(n_{B}^{}) } \, \mathrm{d}n_{B}^{}.
\end{align}
Within the spinodal decomposition region, the chemical potential follows a continuous path from $N$, $I$ to $W$ phase along the isotherms, thus Eqs.~(\ref{eq:free-energy-spinodal}) and (\ref{eq:interface-tension}) can be evaluated in a similar way as \Eq{eq:MC} using the coordinate $\tau$ on the isotherm, to wit:
\begin{align}
  \Delta f(\tau) &= \int_{\tau_{N}}^{\tau} [\mu_{B}^{}(\tau^{\prime}) - \mu_{B}^{\textrm{MC}}] \, \frac{\mathrm{d}n_B(\tau^{\prime})}{\mathrm{d} \tau^{\prime}} \mathrm{d}\tau^{\prime}, \\
  \sigma &= \int_{\tau_{N}}^{\tau_{W}} \sqrt{2 C \, \Delta f(\tau) } \, \frac{\mathrm{d}n_{B}^{}(\tau)}{\mathrm{d} \tau} \mathrm{d}\tau.
\end{align}
In this case, one can also interpret the control parameter $\tau$ as the spatial coordinate within the interface. 

\begin{figure}[htbp]
  \centering
  \includegraphics[width=0.65\columnwidth]{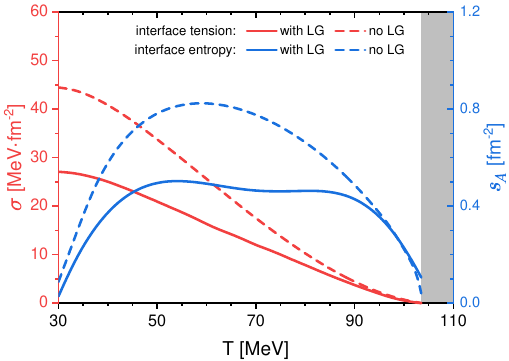}
  \caption{Interface tension $\sigma$ and entropy density $s_A$ as functions of temperature $T$ from $30\,$MeV to $T_{\textrm{CEP}} = 103\,$MeV. The region above $T_{\textrm{CEP}}$ is marked by the gray area, where the QCD phase transition is a crossover and no interface is expected. The case with and without the incorporation of liquid-gas phase transition (labeled by LG) is compared with each other, which are denoted by solid and dashed curves respectively.}\label{fig:surface-tension}
\end{figure}
\begin{figure}[htbp]
  \centering
  \includegraphics[width=0.6\columnwidth]{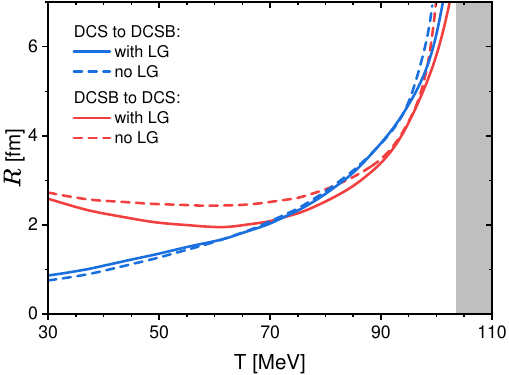}
  \caption{The emergent bubble radius $R$ in dense strong interaction matter as a function of temperature, which is estimated at the two phase boundaries DCS to DCSB (with the decreasing of $T$), and DCSB to DCS (with the increasing of $T$). The region above $T_{\textrm{CEP}} = 103\,$MeV is marked by the gray area, where no bubble formation is expected. The case with and without the incorporation of liquid-gas phase transition (labeled by LG) is compared with each other, which are denoted by solid and dashed curves respectively.}\label{fig:bubble-radius}
\end{figure}

In \Cref{fig:surface-tension}, we show the obtained interface tension as a function of temperature from $T_{\textrm{CEP}} = 103\,$MeV down to 30\,MeV.
The computation is done firstly with no liquid gas transition occurred, i.e. with thermodynamic quantities from the quark sector only. 
In this case, we see that the interface tension increases monotonically towards lower temperature, which can be understood qualitatively as the $\mu_{B}^{}$ gap of the boundaries between the co-exist Nambu and Wigner phase grows larger when $T$ is lowered, as shown in \Cref{fig:phase-diagram}. 
The observed temperature dependence also agrees on the previous findings in Refs.~\cite{Mei:2025ein,Gao:2016hks,Ke:2013wga,Garcia:2013eaa}. 
In particular, the interface tension vanishes towards the CEP at high temperature, and it gradually saturates on the low temperature side. 
The low-temperature limit of the interface tension is found to be around 44\,MeV$\cdot$fm$^{-2}$, which is roughly 3 times of the nucleon binding energy 15\,MeV within a box of the typical length scale 1\,fm for strong interaction. 
In a more realistic case with a liquid-gas transition, the results are qualitatively similar which is different from the situations for the baryon density and for the speed of sound squared. Specifically, the interface tension does have a sizable difference on the low-temperature limit from 44 to 27\,MeV$\cdot$fm$^{-2}$, which follows roughly the magnitude of the volume correction factor $f$ that acts on the quark contribution of thermodynamic quantities within the spinodal region, however its temperature dependence remains unchanged qualitatively. 

The insensitivity of a liquid-gas transition on the interface effect is more clearly seen in the bubble radius measure as proposed originally in Ref.~\cite{Ke:2013wga}. The bubble radius corresponds to the scenarios that the interface between Nambu and Wigner phase acquires a finite curvature, which is possible when the bulk pressure of Nambu and Wigner phase is not equal, as illustrated in the right panel of \Cref{fig:interface}. The bubble radius follows then the Laplacian formula:
\begin{align}
 R = \frac{2 \sigma}{\Delta P},
\end{align}
with $\Delta P$ representing the pressure difference of Nambu and Wigner phase when the interface has a finite curvature, see further details on that in \Cref{app:interface}. 
Two typical measures of $\Delta P$ is the pressure difference on each side of the first-order phase boundary shown in~\Cref{fig:phase-diagram}, which are also referred to as the DCS to DCSB boundary and on the DCSB to DCS boundary respectively in previous literatures. Depending on which phase (Nambu or Wigner) has a larger pressure, there are two types of bubbles whose inner part is composed either by Nambu phase or by Wigner phase. 
The point is that the inner pressure should always be larger than that outside, given that the interface tension is found to be positive, see also the discussion in \Cref{app:interface}. Therefore, according to \Cref{fig:pressure} the inner core of the bubbles is in Nambu phase at the DCSB to DCS phase boundary, while it is in Wigner phase at the DCS to DCSB phase boundary. 

Our results of bubble radius at the two phase boundaries are shown in \Cref{fig:bubble-radius}, where the case with a liquid-gas transition (solid curve) and the case without that (dashed curve) are compared. It is clear that within the given temperature range the overall difference of the radius at DCSB to DCS boundary is no more than 25\%, and the difference is much smaller at DCS to DCSB boundary. In particular, the trends below $T = 50\,$MeV shows that at lower temperature the influence of liquid-gas transition gets smaller, in contrast to being closer to the region where the liquid-gas transition takes place. 
The reason for the insensitivity of liquid-gas transition on the bubble radius relates to the fact that both the interface tension $\sigma$ and the pressure difference $\Delta \tilde{P}$ receives roughly the same rescale factor $f$ from the excluded volume correction, hence their ratio remains almost unchanged implicitly. 
However, it seems that the liquid-gas transition could leave a remnant imprint in the interface entropy
\begin{align}
  s_A  = - \frac{\partial \sigma}{\partial T},
\end{align}
see more on the origin of its definition in \Cref{app:interface}. We show the comparison of $s_A$ as a function of $T$ in \Cref{fig:surface-tension} together with the interface tension. 
In the case without liquid-gas transition, it is found that $s_A$ shows a single-peak structure with the peak position at around $T=60\,$MeV, 
and its value is increasingly small near CEP or towards zero temperature, which follows the general principles of thermodynamic laws. The single-peak structure also agrees with the finding in previous studies~\cite{Gao:2016hks,Ke:2013wga}. 
In the case with liquid-gas transition, $s_A(T)$ shows a double-peak structure instead. This suggests a picture that the liquid-gas transition contributes an enhancement on the interface entropy implicity at low temperature, while its effect are diminished quickly at higher temperature, thus leaving the first peak at lower temperature. Meanwhile, the spinodal decomposition of QCD phase transition produces the second peak at higher temperature, and together we have the double-peak structure. 
We note however that such a double peak structure are not well distinguished considering that their full widths of half maximum has much overlap with each other. Whether such a fine structure is kept in the estimate with an improved knowledge on QCD thermodynamic quantities at high density remains as a task for future works. 
Nevertheless, it is likely that a smooth plateau appears in the range of $T$ from about 50 to 90\,MeV, which is also reflected in the interface tension $\sigma$ with an almost linear trend on $T$ in the respective region. 

We continue to revisit the entropy deficit problem of forming a finite-size bubble during a first-order QCD phase transition. The original motivation of such an investigation is to understand how possible a droplet of nuclear matter or quark-gluon matter can be produced during the phase transition driven by $\mu_B$, as illustrated in Refs.~\cite{Ke:2013wga,Gao:2016hks}: if only considering change of the bulk entropy, the net-entropy deficit of forming a droplet would be negative, see e.g. Refs.~\cite{Blaizot:1999ip,Song:2010bi,Yamazaki:2014psa}, hence the process of hadronization from quarks and gluons seemingly violates the second law of thermodynamics.
This is generally not a complete analysis for a finite-size system where the contribution of interface entropy shall not be neglected. 
In this case, the total entropy of a bubble with volume $V$ and surface area $A$ inside a heat bath with total volume $V_{\textrm{tot.}}$ is given by:
\begin{align}
 S_V + S_A &= V \, s_{V}^{\textrm{in}} + (V_{\textrm{tot.}} - V) \, s_{V}^{\textrm{out}} + A \, s_{A}^{} \nonumber \\
  &= V_{\textrm{tot.}} \, s_{V}^{\textrm{out}} + V (s_{V}^{\textrm{in}} + \tfrac{A}{V} s_{A} - s_{V}^{\textrm{out}}),  \label{eq:total-entropy}
\end{align}
which relates to the curvature of the interface $A/V = 3/R$ in the spherical case. The first term of \Eq{eq:total-entropy} can be understood as the background entropy contribution of the heat bath, hence the net-entropy change when forming a bubble is given by the difference between the effective entropy density $s_V^{\textrm{in}} + \tfrac{A}{V} s_{A}$ of the bubble and the bulk entropy density $s_V^{\textrm{out}}$ of the background. 

\begin{figure}[tbp]
  \centering
  \includegraphics[width=0.48\columnwidth]{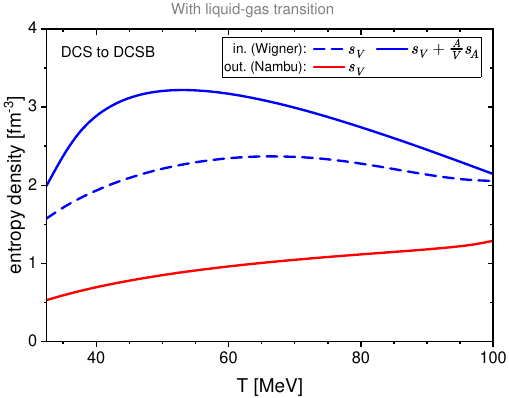}
  \includegraphics[width=0.492\columnwidth]{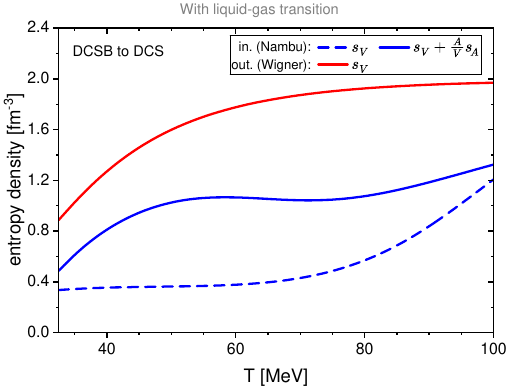} \\[+2ex]
    \includegraphics[width=0.48\columnwidth]{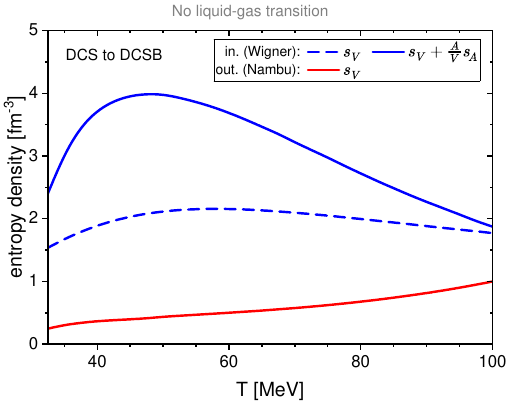}
  \includegraphics[width=0.492\columnwidth]{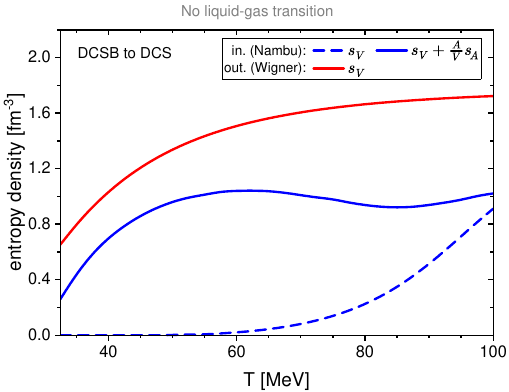}
  \caption{Comparison on the net-entropy density contribution of a bubble ($s_V + \tfrac{A}{V} s_{A}$ inside) and the background ($s_V$ outside), at the DCS to DCSB phase boundary (left) and the DCSB to DCS phase boundary (right). The bulk entropy density for the matter inside the bubble is also displayed as the dashed curves. The case with and without the incorporation of liquid-gas transition are separated in the upper and the lower panel respectively.}\label{fig:entropy}
\end{figure}

In \Cref{fig:entropy}, we compare these two entropy densities at the DCS to DCSB phase boundary (left row) and the DCSB to DCS phase boundary (right row) at different temperatures. 
In all figures, the solid-red curve stands for the background bulk entropy density ($s_V^{\textrm{out}}$) outside the bubble, the dashed-blue stands for the bulk entropy density ($s_V^{\textrm{in}}$) inside the bubble, while the solid-blue curves stands for the sum of bulk and surface contribution of the entropy density ($s_V^{\textrm{in}} + \frac{A}{V} s_A$) inside the bubble. 
Note that according to the pressure at different phases shown in \Cref{fig:pressure}, the matter inside the bubble is in Nambu phase at the DCSB to DCS boundary, while it is in Wigner phase at the DCS to DCSB boundary, see also the discussions in \Cref{app:interface}. 
With these notations, the entropy deficit can be read out from the difference between the blue and the red curves at each of the phase boundary. 
It is found that the bubble formation has a negative entropy deficit at the DCSB to DCS boundary, where the transition process direct to an increasing $T$, while it becomes a positive one at the DCS to DCSB boundary with a direction of decreasing $T$ during phase transition. 
We understand this qualitative difference as in the DCS to DCSB case, 
bubbles with matter in Wigner phase and higher pressure are formed during a fluctuation at high temperature, for example at around $T_{\textrm{CEP}}$. 
As the temperature gets lowered, the radius of such kind of bubbles decreased according to \Cref{fig:bubble-radius}, together with an increase for the pressure difference inside and outside the interface according to \Cref{fig:pressure}. 
Since forming a bubble increases the total entropy of the system, a hadronization process would take place automatically as the temperature decreases. 
While in the DCSB to DCS case, the matter inside the interface is in the Nambu phase which also has a higher pressure than that outside. 
For such kind of bubble, its radius gets enlarged as the temperature rises while its formation leads to a loss on the total entropy, thus the radius would keep increasing drastically and that leads to the DCSB to DCS phase transition. 
Similar comparison is done in the case of no liquid-gas phase transition, see the lower panel of \Cref{fig:entropy}, and the sign of the entropy deficit remains qualitatively the same. 
This is in accordance with the findings in \Cref{fig:bubble-radius}, that the interface properties depends mainly on the phenomenologies within the spinodal region, and the influence of liquid-gas phase transition serves mainly as a background contribution on the bulk thermodynamics for both the Nambu phase and the Wigner phase. 
However, for a quantitative estimate on the entropy deficit, we expect that the result with a liquid-gas phase transition is more realistic, as reflected in the magnitude of the bulk-entropy density of the Nambu phase. This is because the entropy density relates to the baryon density through the thermodynamic relation $\partial s / \partial \mu_B = \partial n_B / \partial T$, and as discussed before the estimation on the Nambu phase baryon density is more reliable when incorporated the liquid-gas phase transition, which corresponds to a relatively large entropy density.

In the last part of this Section, we investigate from a new perspective on the stability of bubble formation, which goes deeper from the entropy deficit analysis. 
The idea follows that the entropy deficit analysis essentially aims at understanding the thermodynamic stability of bubble formation. For a self-consistent judgement on the bubble stability, here we refer to the compressibility of $\kappa = \partial^2 F_{V+A}^{} / \partial V^2$ defined by the total free energy $F_{V+A}$ of finite-size bubble including both the bulk and the interface contributions. 
This directly follows the equilibrium condition in \Eq{eq:bubble-EoM} with a consideration on the volume fluctuations, that a positive $\kappa$ is required to keep the bubble stable. 
Here we shall consider the case that the heat bath outside the bubble keeps as an isobaric background, whose temperature as well as the temperature inside the bubble stays constant. 
With \Eq{eq:first-thermo-law}, the compressibility in this case follows: 
\begin{align}
  \kappa = \frac{\partial^2 F_{V+A}^{} }{ \partial V^2 } = -\frac{\partial P_{\textrm{in}}}{\partial V} + \sigma \frac{\mathrm{d}^2 A}{\mathrm{d} V^2}. \label{eq:bubble-stability}
\end{align}
The first term represents the bulk compressibility of the matter inside the bubble, with $P_{\textrm{in}}$ the inner pressure. 
As illustrated above, $P_{\textrm{in}}$ matches the Wigner phase for the DCS to DCSB case, while it matches the Nambu phase for the DCSB to DCS case. 
Due to the net-baryon number conservation inside the nuclear bubble, the bulk compressibility can be further specified as:
\begin{align}
  \kappa_V = -\biggl(\frac{\partial P_{\textrm{in}}}{\partial V}\biggr)_{T,N_{B,\textrm{in}}} = \frac{n_{B,\textrm{in}}^{2}}{N_{B,\textrm{in}}} \biggl(\frac{\partial P_{\textrm{in}}}{\partial n_{B,\textrm{in}}^{}}\biggr)_{T} = \frac{n_{B,\textrm{in}}^{2}}{V} \biggl(\frac{\partial n_{B,\textrm{in}}^{}}{\partial \mu_{B}^{}}\biggr)_{T}^{-1},  \label{eq:bulk-compress}
\end{align}
with $N_{B,\textrm{in}} = n_{B,\textrm{in}}^{} V$ the net-baryon number conserved inside the bubble. 
The second term in \Eq{eq:bubble-stability} is in general negative, as the interface tension tends to shrink the total surface area of the bubble. For a spherical bubble, one can show that:
\begin{align}
  \sigma \frac{\mathrm{d}^2 A}{\mathrm{d} V^2} = -\frac{\sigma}{2\pi R^4} = -\frac{\Delta P}{3V}, \label{eq:interface-compress}
\end{align}
where we have taken \Eq{eq:bubble-estimate} into account for the interface tension $\sigma$. 
The stability condition for the bubble can then be understood as the competition between the bulk compressibility and the interface effect.

\begin{figure}[tbp]
  \centering
  \includegraphics[width=0.49\columnwidth]{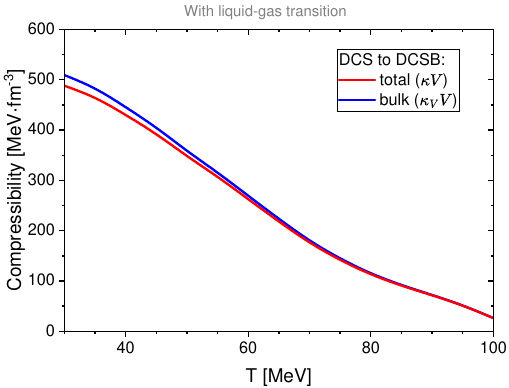}
  \includegraphics[width=0.485\columnwidth]{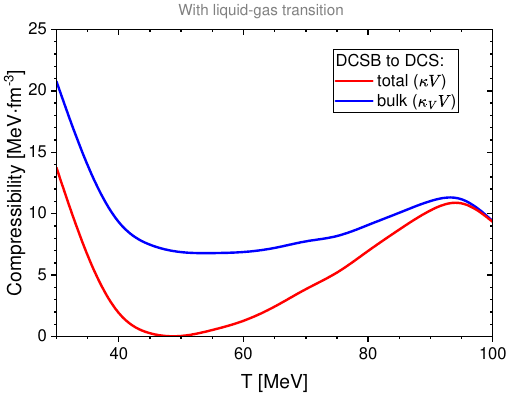}
  \caption{Total compressibility $\kappa$ in \Eq{eq:bubble-stability} versus bulk compressibility $\kappa_V$ in \Eq{eq:bulk-compress}, for the nuclear bubble formed at the DCS to DCSB phase boundary (left) and at the DCSB to DCS boundary (right). A positive $\kappa$ is required to keep the formed bubble stable under volume fluctuations. Both results are scaled by the bubble volume $V=\frac{4}{3} \pi R^3$, with the respective radius $R$ given in \Cref{fig:bubble-radius} at different temperatures.}\label{fig:compressibility}
\end{figure}

In \Cref{fig:compressibility}, we show the bulk compressibility $\kappa_V$ calculated with the $n_B$ in the left panel of \Cref{fig:net-density}, versus the total compressibility $\kappa$ determined by Eqs.~(\ref{eq:bubble-stability}) and (\ref{eq:interface-compress}) with inputs of the interface tension in \Cref{fig:surface-tension} and the bubble radius in \Cref{fig:bubble-radius}, for the case at the DCS to DCSB boundary and the case at the DCSB to DCS boundary. 
At DCS to DCSB boundary, The inner core is in Wigner phase and the bulk compressibility has a much larger magnitude than the interface compressibility, hence resulted in a positive-definite total compressibility of the bubble formation. This implies that quark droplets at the DCS to DCSB boundary is possible to emerge for all temperatures below $T_{\textrm{CEP}}$, which supports the emergence of QCD supercooling phenomenon in the coexistence region of Nambu and Wigner phase. 
At the DCSB to DCS boundary, the compressibility becomes relatively small and comparable with the interface compressibility, considering that in the inner core of the bubble is now composed of Nambu phase. 
For the total compressibility, we still arrive at a (semi-)positive result for all temperature, hence it is also likely for the emergence of overheating phenomenon in QCD phase transition. 
However, we also see that the total compressibility is nearly vanishing at around $T=50\,$MeV, which is also where the minimum of bulk compressibility is located. 
We understand that due to the suppression of bulk compressibility around this temperature, the inner core of the bubble becomes less sustainable to compete with the interface tension, hence the stability of the forming an overheating bubble becomes lower. 
Meanwhile, we see that the enhancement of bulk compressibility below $T=50\,$MeV is due to the presence of liquid-gas phase transition before reaching the QCD phase transition boundary, which in turn increases the stability of overheated bubbles. Just as indicated in the sub-structure in the interface entropy \Cref{fig:surface-tension}, this implies that the imprint of liquid-gas phase transition may still be found in the regime of QCD phase transition through the high-order fluctuations, note that the compressibility refers to the density fluctuation according to \Eq{eq:bulk-compress}, while the sub-structure of $s_A(T)$ discussed before is essentially about the (interface) heat capacity $c_{A} \propto \frac{\partial s_A}{\partial T}$ which refers to thermal fluctuations. 
In all, we verified with the compressibility analysis that both the supercooling and overheating phenomena of QCD phase transition can play a role to incorporate the inhomogeneous structures in the dense strong interaction matter. The possible impact of these phenomena on the observational signals in nuclear experiments and astrophysical objects will be investigated in the near future.

\section{Conclusion and discussion}\label{sec:summary}

In this work, we revisited the first-order QCD phase transition and the thermodynamic observables in the respective region using the Dyson-Schwinger equations approach. 
An individual solution branch is found in the quark gap equation which represents an intermediate phase within the co-exist region of Nambu and Wigner phase, for all temperatures below that of the critical end-point. 
This solution  verifies that the spinodal decomposition appears as a general phenomenon in first-order QCD phase transition, both in the multi-phase structure of the quark propagators which reflects the microscopic dynamics of chiral symmetry breaking in the quantum equation of motion, and also in the macroscopic features of thermodynamics quantities, including the chiral condensate, Polyakov loops and  the equation of state.

For a more realistic description on the Nambu phase near the liquid-gas phase transition, we adopt a mixing approach between the mean-field Walecka model and the DSE result on the QCD equation of state at high baryon density, by taking into account the excluded volume correction between hadrons and quarks. The result implies that the implicit influence of liquid-gas phase transition plays a crucial role in the thermodynamic properties, which stiffens the equation of state regarding the qualitative changes in the baryon density and the speed of sound. While on the other hand, the spinodal decomposition in the QCD phase transition generally provides a softening on the equation of state. 

By further constructing an inhomogeneous distribution of nuclear density, it is allowed for an improved prediction on the interface tension and the nuclear bubble radius with temperature dependencies after making a complete analysis of the dynamics during the QCD first order phase transition based on the stationary condition of the free energy. 
It is found through comparison that the influence of liquid-gas phase transition on the interface effect becomes less obvious than what is shown in the bulk thermodynamics at low temperature, while it is still possible that it leaves an imprint in the high-order fluctuations during the QCD phase transition. 
The interface tension is also adopted for an re-investigation on the entropy deficit problem of hadronisation process. Similar to the estimates in previous studies, it is found that the process shows qualitatively different signatures at DCS to DCSB boundary and at DCSB to DCS boundary, while it is also found that here the influence of liquid-gas phase transition is mild. 
In addition, the stability of the formed nuclear bubble is checked via the net-compressibility of the bubble. The net-compressibility stems from the second derivative of the total free energy of a finite-size system, which as a novel quantity suggests that both supercooling and overheating phenomena might take place which are responsible for the inhomogeneous phase structures and droplet formations in the dense strong interaction matter. 

We note that the present study focus only on the bulk inhomogeneity in association to the dynamics of the first order phase transition, and there may also exist  microscopic  inhomogeneity,   which are referred to as the moat regime, inhomogeneous condensates and color superconductivity. Their impacts on the thermodynamics can be non-trivial and will be further studied. 
Nevertheless, the improved results on thermodynamic observables for dense nuclear matter can be helpful as inputs for further combination with heavy-ion physics or astrophysics research.

\acknowledgments

We thank the members of the fQCD collaboration~\cite{fQCD} for discussions.
YL and YXL are supported by the National Science Foundation of China under Grants  No.~12175007 and No.~12247107. 
FG is supported by the National  Science Foundation of China under Grants  No.~12305134. 

\newpage

\appendix
\textbf{\large{Appendix}}
\section{Homotopy method for the numerical iterations on the gap equation}\label{app:homo-init-iter}

Practically, the numerical iterative computation on the quark gap equation in \Cref{fig:quark-DSE} requires an input of the initial quark propagator $G_q^{\textrm{init.}}$, which is a function of Matsubara frequency and spatial momentum. In previous studies, the Nambu($N$)/Wigner($W$) solution branch is found typically when such an initial propagator takes a large/small mass with respect to the vacuum mass scale $M_q^{\textrm{vac.}} \approx 350\,$MeV, see e.g.~\cite{Gunkel:2019xnh,Gao:2015kea}. 
However, towards the low $T$ and high $\mu_{B}^{}$ region, the computation of gap equation turns out to be much complicated with a sensitivity on the choice of the initial propagator for the iteration process. 
Inspired by the homotopy method~\cite{Zheng:2023tbv,Wang:2012me}, we come up with an improvement on the initial propagator based on a linear combination of $N$ and $W$ solutions:
\begin{align}\label{eq:inital-prop-homo}
 G_q^{\textrm{init.}}(p;\eta) = \eta G_q^{N}(p) + (1-\eta) G_q^{W}(p), 
\end{align}
with $\eta$ the homotopy parameter which is a real constant. \Cref{eq:inital-prop-homo} allows for a continuous modification on the initial propagator from the chiral symmetry breaking solution to the chiral symmetric solution, which turns out to be helpful both for finding all possible solutions in the gap equation and for testing the numerical sensitivity of the initial condition. 
It should be emphasised that $\eta$ is only an auxiliary variable for identifying different solution branches and shall not be understood as an input parameter of the theory. 

\begin{figure}[bp]
  \centering
  \includegraphics[width=0.48\columnwidth]{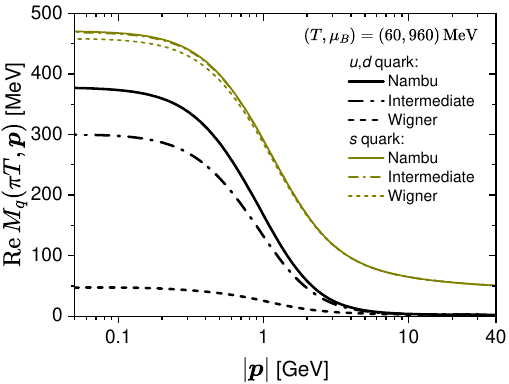}
  \includegraphics[width=0.48\columnwidth]{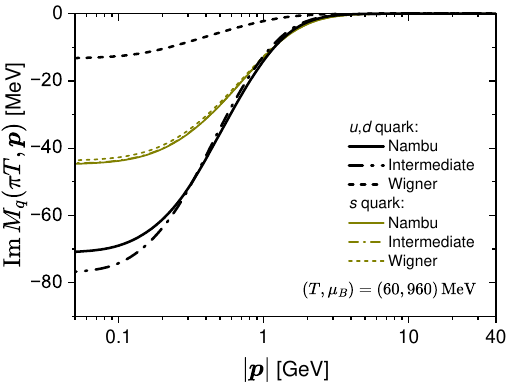}
  \caption{Spatial momentum ($\spatial{p}$) dependence of the $u,d$ and $s$ quark mass functions at $\omega_p = \pi T$ for the multi-phases $N$, $I$ and $W$, including their real and imaginary parts, at $(T,\mu_{B}^{})=(60,960)\,$MeV. The mass at $\spatial{p}=0$ is in match with \Cref{fig:mass-branches-T}. }\label{fig:app-mass-spatial-mom}
\end{figure}
As discussed in the main text, we verified for all possible $\eta \in (0,1)$ numerically that there is one unique solution $G_{q}^{I}$ found other than $G_{q}^{N}$ and $G_{q}^{W}$. 
Specifically, the final solution is distinguished by two critical values $\eta_N$ and $\eta_W$ (with $0 < \eta_W < \eta_N < 1$): the Wigner, intermediate and Nambu solutions are obtained for $\eta \in [0,\eta_W)$, $(\eta_W,\eta_N)$ and $(\eta_N, 1]$, respectively. 
This also means that the solutions for quark gap equation are insensitive with respect to in the initial condition as long as $\eta$ stays within one of the ranges given above.
For completeness, we further demonstrate here the spatial-momentum ($\spatial{p}$) dependence of the $u,d$ and $s$ quark mass function at frequency $\pi T$, for better understanding on the multi-phase structure for readers. We choose the case at $(T,\mu_{B}^{})=(60,960)\,$MeV, and the results are shown in \Cref{fig:app-mass-spatial-mom}, and the mass values at $\spatial{p}=0$ are in match with \Cref{fig:mass-branches-T}. 
Especially, the results imply that the strange quark mass barely changes   in the $T$ and $\mu_{B}^{}$ region that we focus on, as discussed around \Eq{eq:redcued-cond} which makes the reduced condensate a good measure of the chiral symmetry breaking for $u$ and $d$ quarks.

In fact, for other values of $\eta$, typically when $\eta < 0$, the negative Nambu solution can also be obtained using \Eq{eq:inital-prop-homo}, which has $\Re M_q < 0$ and it results in a negative chiral condensate. However, the investigation on the solution branches with a negative condensate is beyond the scope of this work. For related discussions, please refer to \cite{Wang:2012me}.

\section{Interface properties and thermodynamics}\label{app:interface}

Here we briefly review some of the key connections between the interface properties and thermodynamic quantities. The related discussions can also be found in~\cite{Ke:2013wga,Gao:2016hks}. 
Considering a deformation of the interface with the variance on temperature $T$, volume ($V$) and surface area ($A$), 
the change of total free energy is contributed from both the bulk part ($V$) and the interface part ($A$):
\begin{align}
  \mathrm{d}F_{V+A}^{} = \mathrm{d}F_{V}^{} + \mathrm{d}F_{A}^{}.
\end{align}
Following the thermodynamic laws, the bulk part reads:
\begin{align}
  \mathrm{d}F_{V}^{}(T,V) =  -\Delta S_V^{} \mathrm{d}T - \Delta P \mathrm{d}V,
\end{align}
with $\Delta$ representing the difference of a thermodynamic quantity inside and outside the bubble, $P$ the pressure, and $S_{V}^{}$ the bulk entropy which is independent of $A$. 
Meanwhile, the surface part is determined by: 
\begin{align}
  \mathrm{d}F_{A}^{}(T,A) = -S_A^{} \mathrm{d}T + \sigma \mathrm{d}A,
\end{align}
with $S_{A}^{}$ the interface entropy and $\sigma$ the interface tension as discussed above. 
This entails the thermodynamic law for the system with an interface as:
\begin{align}
  \mathrm{d}F_{V+A}^{}(T,V,A) = -(\Delta S_V + S_A) \mathrm{d}T - \Delta P \mathrm{d}V + \sigma \mathrm{d}A. \label{eq:first-thermo-law}
\end{align}
This immediately allows for an evaluation on the interface entropy density $s_A$, which is just the temperature susceptibility of $\sigma$:
\begin{align}
  s_A \coloneqq \bigl( \frac{\partial S_A}{\partial A} \bigr)_V = \frac{\partial (\Delta S_V + S_A)}{\partial A} = - \frac{\partial \sigma}{\partial T}.
\end{align}

On the other hand, the equilibrium condition $\mathrm{d}F_{V+A}^{}=0$ for a given temperature yields: 
\begin{align}
  \Delta P = \sigma \bigl(\frac{\mathrm{d}A}{\mathrm{d}V}\bigr)_{T}.   \label{eq:bubble-EoM}
\end{align}
This equation implies that given the pressure difference of the coexist Nambu and Wigner phases, which is in general finite within the first-order phase boundaries, it is possible for the formation of nuclear bubbles its geometric properties determined by the strength of interface tension $\sigma$. 
In particular for a spherical bubble, the bubble radius $R$ can be estimated as $\mathrm{d}V$ and $\mathrm{d}A$ are matched with the radius as:
\begin{align}
  \frac{\mathrm{d}V}{\mathrm{d}A} = \frac{\mathrm{d}V(R)/\mathrm{d}R}{\mathrm{d}A(R)/\mathrm{d}R} = \frac{R}{2}, 
\end{align}
thus we have:
\begin{align}\label{eq:bubble-estimate}
  R = \frac{2\sigma}{\Delta P},
\end{align}
which is structurally similar to the Laplacian formula for descring the surface tension of a liquid droplet. 
Along the first-order phase boundary, i.e. boundaries of the shadowed area in \Cref{fig:phase-diagram}, the pressure difference $\Delta P$ between the Nambu phase and the Wigner phase changes with respect to the chemical potential, and the bubble radius changes accordingly. 
In the main text, we have specifically focused on the two phase boundaries shown in \Cref{fig:phase-diagram}. 
With a positive interface tension, \Eq{eq:bubble-EoM} implies that the pressure inside the bubble is always larger than that outside. 
Hence according to \Cref{fig:pressure}, the matter inside the interface is in Nambu phase for the DCSB to DCS case, which has a lower baryon number density than that outside and plays the role of a seed for the DCSB phase. 
While for the DCS to DCSB case, the bubble is composed of the Wigner phase with a higher baryon number density and it functions as the infant of a hadron.

\newpage

\bibliographystyle{JHEP}
\bibliography{interface_v2.bib}

\end{document}